\documentstyle[preprint,aps,epsf,amsmath,amssymb,graphicx]{revtex}
\begin{document}

\preprint{\vbox{ \begin{flushright}
                   THEF-NIJM 01.01 \\
                   KVI-1522
                 \end{flushright}
               }
         }
\draft
\tighten

\title{Normalization of neutron-proton differential cross sections}

\author{M.C.M. Rentmeester,$^a$ R.G.E. Timmermans,$^b$ J.J. de Swart$^a$}

\address{$^a$Institute for Theoretical Physics, University of Nijmegen,
         P.O. Box 9010, \\ 6500 GL Nijmegen, The Netherlands}
\address{$^b$Theory group, KVI, University of Groningen,
         Zernikelaan 25, \\ 9747 AA Groningen, The Netherlands}

\maketitle

\begin{abstract}
The $np$ differential cross section below 350 MeV neutron laboratory
energy is studied using the energy-dependent Nijmegen partial-wave
analysis PWA93.
We analyze in detail three experiments, performed at LAMPF, at TRIUMF,
and at the TSL facility in Uppsala.
The issue of normalization of $np$ cross sections is discussed,
where we distinguish between measured and calculated normalizations.
This analysis leads to improved treatments of the LAMPF and TRIUMF
data over PWA93.
It is shown that the LAMPF and TRIUMF data at $T_{lab}$ = 212 MeV are
in good agreement with PWA93, that the LAMPF data at $T_{lab}$ = 162
MeV is also in good agreement with PWA93, but that
the Uppsala data at 162 MeV is in strong disagreement with PWA93.
The reason for the disagreement is, almost certainly, a systematic
flaw in the slope of the Uppsala data.
\end{abstract}
\pacs{PACS numbers: 11.80.Et, 13.75.Cs, 13.75.Gx, 21.30.-x}

\section{Introduction}
The neutron-proton ($np$) differential cross section at neutron laboratory
energies below 350 MeV has been a topic of frequent investigations.
The relevant $np$ data base can be found in NN-OnLine~\cite{NNOnL}
and in SAID~\cite{SAID}.
One reason for the special interest in this cross section has been the
suggestion made by G.F. Chew~\cite{Che58} in 1958 that the pion-nucleon
coupling constant could be determined from the backward $np$ data.

\begin{figure}[t]
\begin{center}
\includegraphics{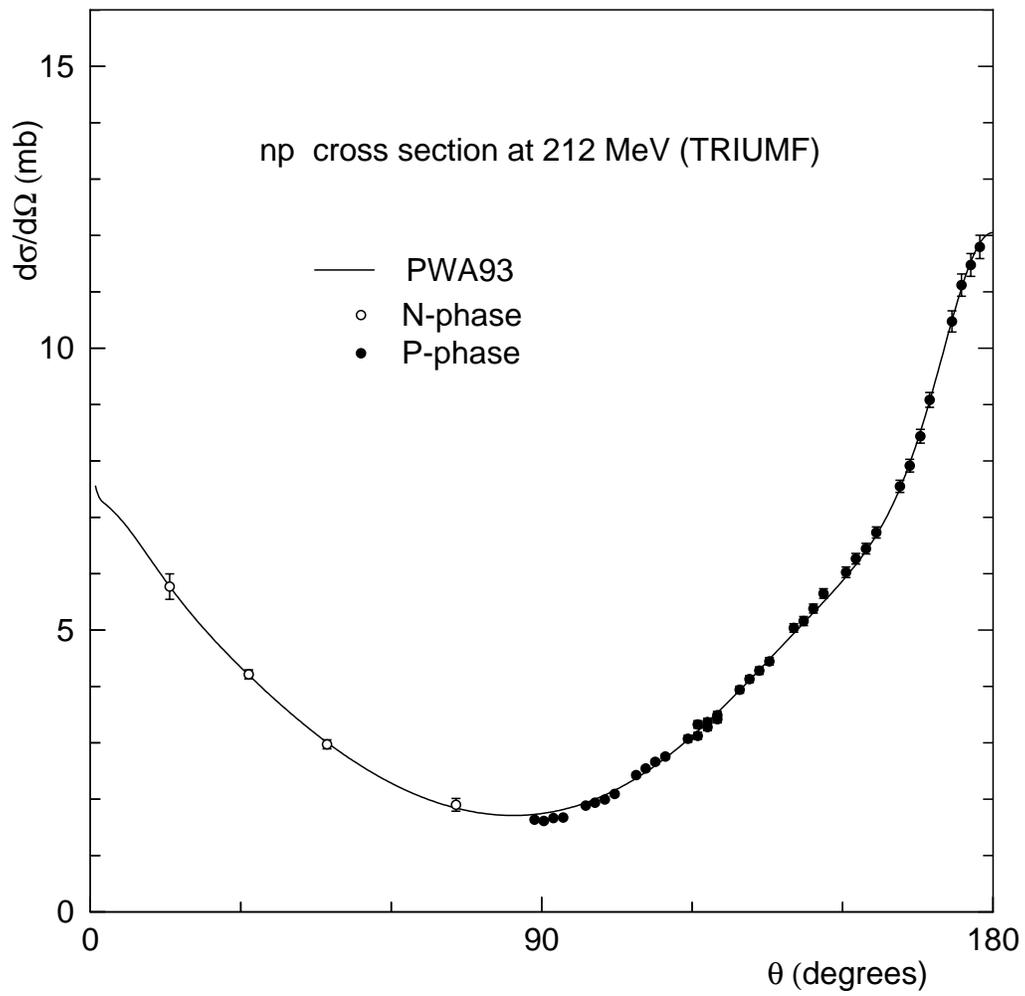}
\end{center}
\caption{The $np$ differential cross section at 212 MeV.}
\label{fig:1}
\end{figure}

In Fig.\ref{fig:1} we show the $np$ differential cross section at
$T_{lab}$ = 212 MeV as predicted by the energy-dependent Nijmegen
partial-wave analysis PWA93~\cite{Sto93}, together with absolutely
normalized TRIUMF data~\cite{Kee82}. The most distinctive features of
this cross section are the forward peak, due to the destructive interference
between neutral-pion exchange and the rest of the amplitude, and the
backward peak, similarly due to the destructive interference between
charged-pion exchange and the rest of the amplitude.

The measurement of an $np$ differential cross section is notoriously
difficult. Especially the determination of the correct normalization
often poses problems.
These cross sections are most easily measured in the backward direction,
because the then forward-going recoil protons have in that case
a relatively high energy.
In order to cover a larger angular region, often several different
settings of the detection apparatus are used. The differential cross
section is then measured in different, sometimes overlapping, angular
regions. The $np$ data base~\cite{NNOnL,SAID,Sto93} with $T_{lab}<350$ MeV
contains a number of such data sets~\cite{Kee82,Bon78,Eri95,Rah97,Rah98}.

In this paper we will analyze mainly the $np$ backward cross section data,
where we will focus in particular on the way the data sets are normalized.
We will study how the sets in different angular regions are normalized
`relative' to each other, and how the `absolute' normalizations of the
various cross sections are obtained.
Sometimes these normalizations (relative or absolute) are actually
measured~\cite{Kee82,Bon78}, we speak then of `measured' normalizations,
but in other cases these normalizations are determined via some model
calculation~\cite{Eri95,Rah97,Rah98}, we speak then of `calculated'
normalizations. This has consequences for the way the data sets
should be treated in partial-wave analyses (PWA's) of $np$ scattering
data~\cite{NNOnL,SAID,Sto93,Klo94,Ren01,Arn00}.
`Measured' normalizations are experimental data with an
associated error and must be included in the data base.
`Calculated' normalizations are {\em not} experimental data and
therefore should {\em not} be included in the data base.
In PWA's calculated normalizations should be floated.
Therefore, it would be helpful if the experimentalists would clearly
indicate if the normalizations they use are actually measured or
merely calculated.
`Calculated' normalizations depend on the model used to determine these
normalizations. Energy-dependent PWA's are especially well suited to
determine these calculated normalizations.
At this moment the Nijmegen partial-wave analysis PWA93~\cite{NNOnL,Sto93}
is probably one of the best tools to obtain them.
Since we use PWA93 we are restricted to data sets with $T_{lab}<350$ MeV.

We will study in this paper in detail three different $np$ experiments.
One of the most important data sets was measured at LAMPF~\cite{Bon78}.
Next, there is an experiment done at TRIUMF~\cite{Kee82}, and another
one performed at the TSL facility in Uppsala~\cite{Eri95,Rah97,Rah98}.
We will concentrate our efforts on two neutron laboratory energies:
162 MeV and about 212 MeV. The reason for this is that at 162 MeV
there are two data sets (LAMPF and Uppsala) available, while at 211.5
MeV there is a set from LAMPF and at 212 MeV a set from TRIUMF.
These data sets can then be compared via our PWA93.
The LAMPF and TRIUMF data sets were included in PWA93 and in general
both sets were fitted very well in this partial-wave analysis. However,
it turned out to be impossible for us to get a satisfactory fit to the
later measured Uppsala data set, despite large efforts~\cite{Ren97,Ren98}
from our side.
The reason resides in systematic discrepancies between the Uppsala
data and PWA93, which represents in a kind of averaged way the world
$np$ data base.

This paper is neither a study of all $np$ data, nor a study of all
$np$ differential cross sections. It is a study of the manner in which
differential cross sections could be normalized. We consider three
different cases, each with its own specific way of normalization.
We hope to convince the reader of the high accuracy that can be
reached by PWA's when normalizing $np$ differential cross sections.

In Section~\ref{EXP} we will discuss the various data sets and the way
in which they were treated in or with the Nijmegen PWA93.
We will allow only the absolute normalization to be adjusted to the data.
In Section~\ref{REL} we will allow also the relative normalization
between the various subsets to be adjusted.
This leads to significantly improved fits in all cases considered.
Finally, in Section~\ref{CON} we will summarize our conclusions.

\section{Experimental data and absolute normalization} \label{EXP}
For each differential cross section the absolute normalization is
denoted by $N_{abs}$ and its error by $dN_{abs}$. When this is a
{\em measured} normalization these numbers are used in the PWA.
In PWA93~\cite{Sto93} each differential cross section contributes
to $\chi^2$ as
\begin{equation}
    \chi^2 = \chi^2(data) + \chi^2(norm) = \sum_i
    \left[\frac{N_{abs}\:\sigma(\theta_i,exp) - \sigma(\theta_i,pwa)}
    {N_{abs}\: d\sigma(\theta_i,exp)}\right]^2 +
    \left[\frac{N_{abs}-1}{dN_{abs}}\right]^2 \ ,
\end{equation}
where $\sigma(\theta_i,exp) \pm d\sigma(\theta_i,exp)$ is the value
of the measured differential cross section with its error.
We need to sum over all data points $i$ of the data set.
In case the absolute normalization has not been measured,
but is just a calculated or even a floated normalization,
this normalization should not contribute to $\chi^2$. This is
easily achieved in practice by assigning a very large value to $dN_{abs}$.

\subsection{The LAMPF data}
At LAMPF~\cite{Bon78} a neutron beam with a continuous spectrum was
obtained by passing an 800 MeV proton beam through an aluminum
production target.
This neutron beam was directed on a liquid-hydrogen scattering
target. The recoiling protons were detected with a magnetic multiwire
proportional chamber (MWPC) spectrometer.
The proton laboratory angle varied from 0$^\circ$ to about 30$^\circ$.
The experimental data was divided into 30 MeV/c wide bins for the
laboratory momentum of the incident neutrons. These bins were centered
in 30 MeV/c steps ranging from 575 MeV/c to 1385 MeV/c. An
extra bin was centered at 1429 MeV/c, which contains the data with
$p_{lab}>$ 1400 MeV/c.
This results in a total of 29 energy bins from 162 MeV to about 700 MeV,
where 11 of these bins have a central laboratory energy below 350 MeV.

The absolute normalizations of the differential cross sections at
the lowest 7 energies were floated.
For neutron energies larger than 280 MeV the differential cross sections
were absolutely normalized with the help of the simultaneously detected
deuterons from the reaction $n+p\rightarrow d+\pi^0$.
Via charge independence this reaction is related to the well-known
reaction $p+p\rightarrow d+\pi^+$.
The overall accuracy of this measured normalization is in the
range of 5 to 10\%.

Because we want to compare the data with PWA93 we can use only 11 of
these 29 cross sections. The results are presented in Table~\ref{tab:1}.
In the analysis for PWA93 the data set at 194.5 MeV consisting of 42 data
had $\chi^2$ = 76. This $\chi^2$ is too large, indicating a more than
3 standard deviation (s.d.) discrepancy.
Also the data point at 131.51$^{\circ}$ in the 344.5 MeV set had a
too high $\chi^2$.
Therefore, the data set at 194.5 MeV and the single data point at 344.5
MeV and 131.51$^{\circ}$ were not included in the final data base for PWA93.
In the analysis presented here we returned this data in first instance
back into our data base. When the analysis was done we noticed that the
$\chi^2$ for the data point at 344.5 MeV and 131.51$^{\circ}$ was still
too large. This data point was therefore again removed from the data set.

The 11 differential cross sections contain in total 650 data.
After removing the data set at 194.5 MeV and the single data point
at 344.5 MeV the data set for PWA93 consisted of 607 points. For
these 607 data we expected $\langle\chi^2\rangle = 597\pm 35$.
We reached $\chi^2 = 630$, which indicates a good fit.
In PWA93 we could determine the absolute normalization of each of
the differential cross sections with an accuracy of 0.7\% or better.
For the sets with $T_{lab} >$ 280 MeV the accuracy was even 0.5\%
or better; see Table~\ref{tab:1}.
This is about a factor of 10 to 20 better than the absolute normalizations
measured with the help of the reaction $n+p\rightarrow d+\pi^0$.
Most of the uncertainties in the calculated absolute normalizations of
these data sets comes from the statistical errors in the individual
data points. The uncertainty due to the error in our knowledge of
the total cross section is much less.

\begin{figure}[t]
\begin{center}
\includegraphics{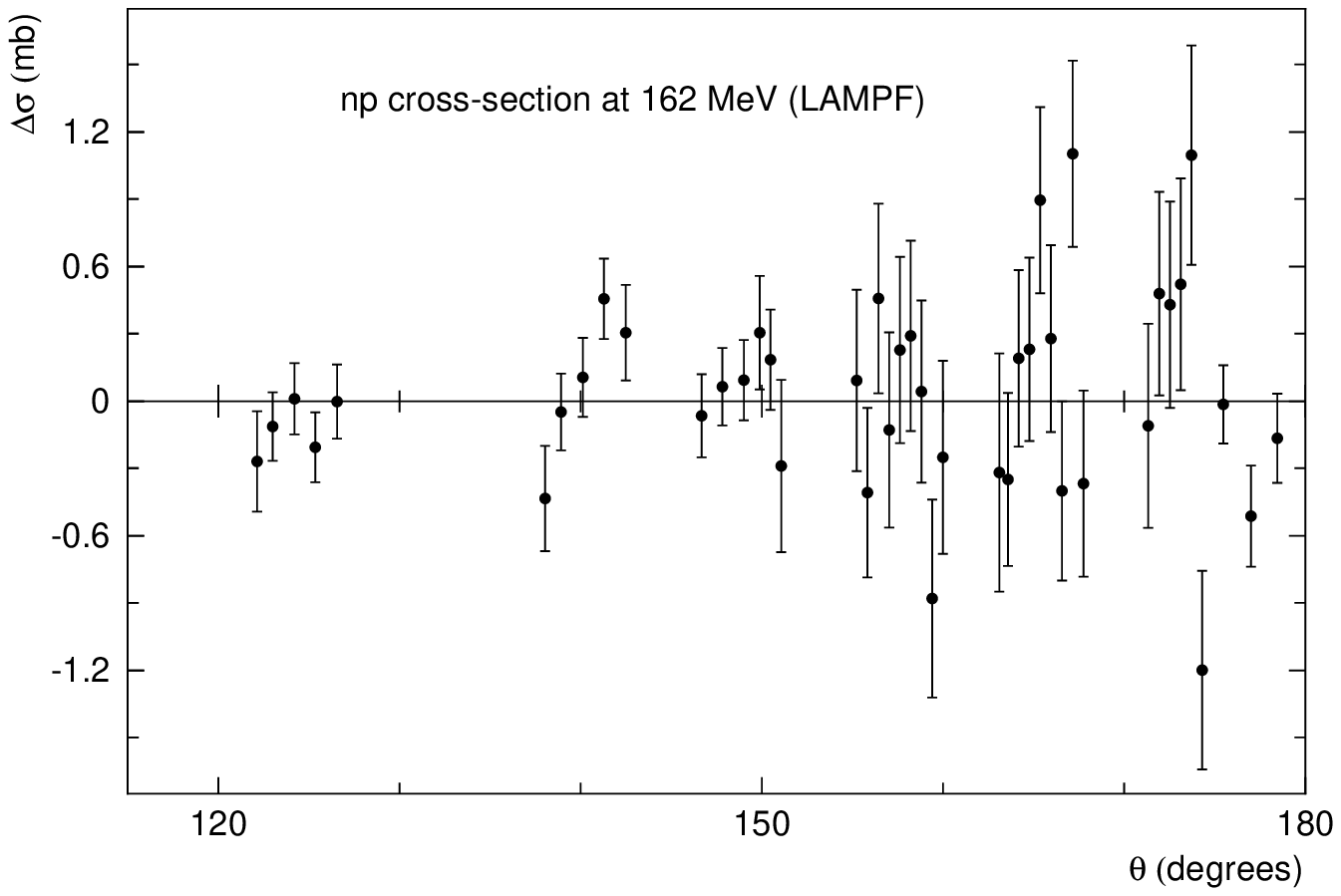}
\end{center}
\caption{The difference $\Delta\sigma(\theta)$ between the absolutely
         normalized LAMPF data and PWA93 at 162 MeV.}
\label{fig:2}
\end{figure}

\begin{figure}[t]
\begin{center}
\includegraphics{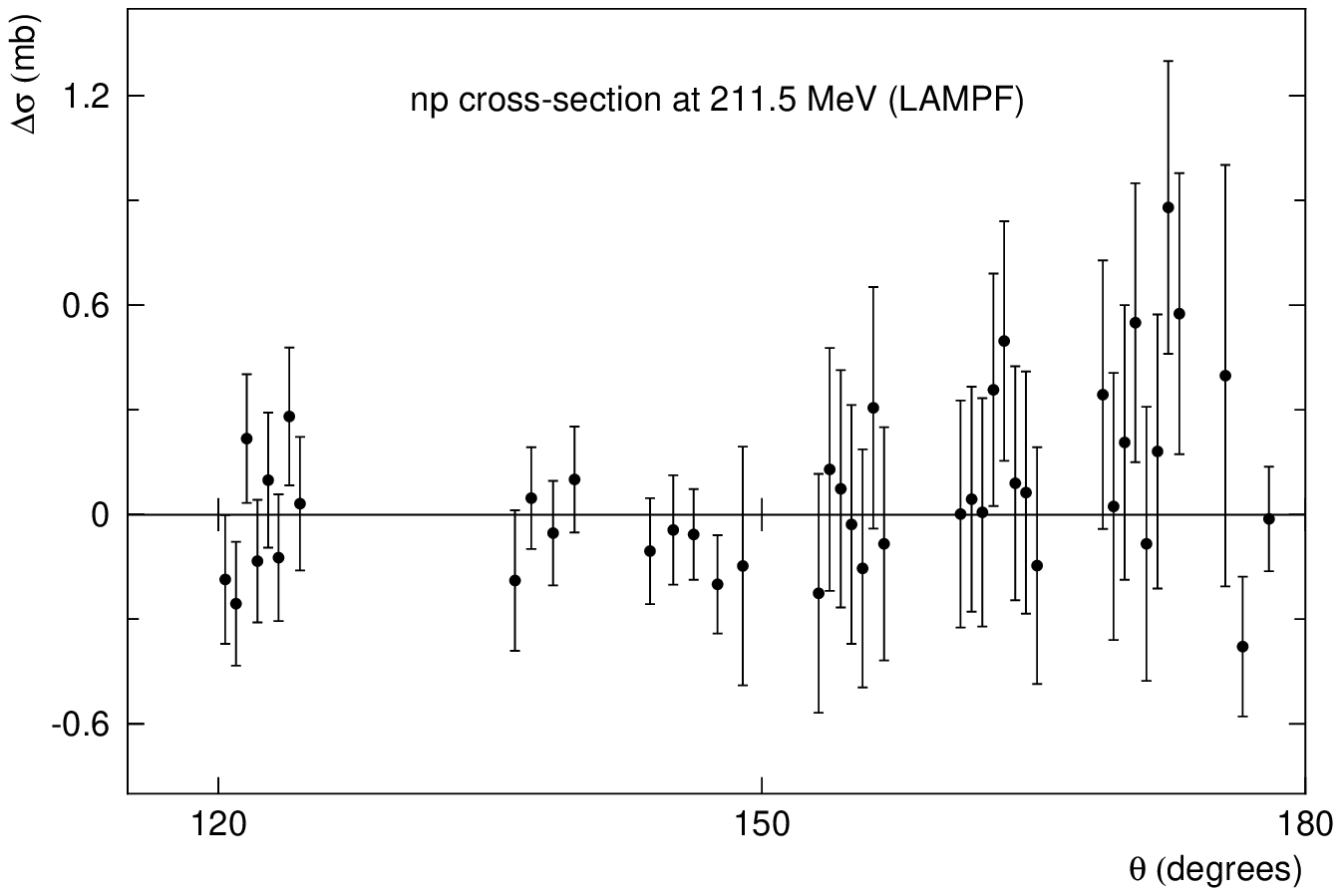}
\end{center}
\caption{The difference $\Delta\sigma(\theta)$ between the absolutely
         normalized LAMPF data and PWA93 at 211.5 MeV.}
\label{fig:3}
\end{figure}

In Fig.~\ref{fig:2} we show the difference
$\Delta\sigma = N_{abs}\sigma(exp) - \sigma(pwa)$
between the absolutely normalized LAMPF data at 162.0 MeV and PWA93.
In Fig.~\ref{fig:3} the same is done for the 211.5 MeV data.
Several things can be noticed in these Figures.
First of all, we see that the data sets can be divided into
7 non-overlapping subsets, each covering
a small ($\simeq 4.5^\circ$) angular region.
We number these subsets 1 to 7, starting at the smallest neutron angle.
Because we have for each energy only 1 overall normalization available
we cannot optimally normalize these individual subsets. For the 162
MeV data we see that set 1 (5 points around 124$^{\circ}$) and set 7
(3 points around 177$^{\circ}$) would prefer a larger normalization.
When we decrease the normalization of set 6 (6 points around $173^\circ$),
5 points from this set obtain a more satisfactory agreement with PWA93,
but the point at 174.32$^\circ$ will fit even worse.
In our final fit the latter point will be more than 3 s.d. off.
We will therefore remove this point already now from our data set.
The fit with the remaining 42 data at 162.0 MeV has
$\chi^2 = 56$ with $N_{abs} = 1.090(7)$.

\subsection{The TRIUMF data}
At the TRIUMF fast-neutron facility $np$ scattering cross
sections were measured between 200 MeV and 500 MeV~\cite{Kee82}.
The neutrons were produced in a liquid-deuterium production target
by the inclusive reaction $p+d\rightarrow n+X$.
The quasi-elastic neutrons were used for the analysis.
The neutrons produced in the forward direction were
collimated on a liquid-hydrogen scattering target.

\begin{figure}[t]
\begin{center}
\includegraphics{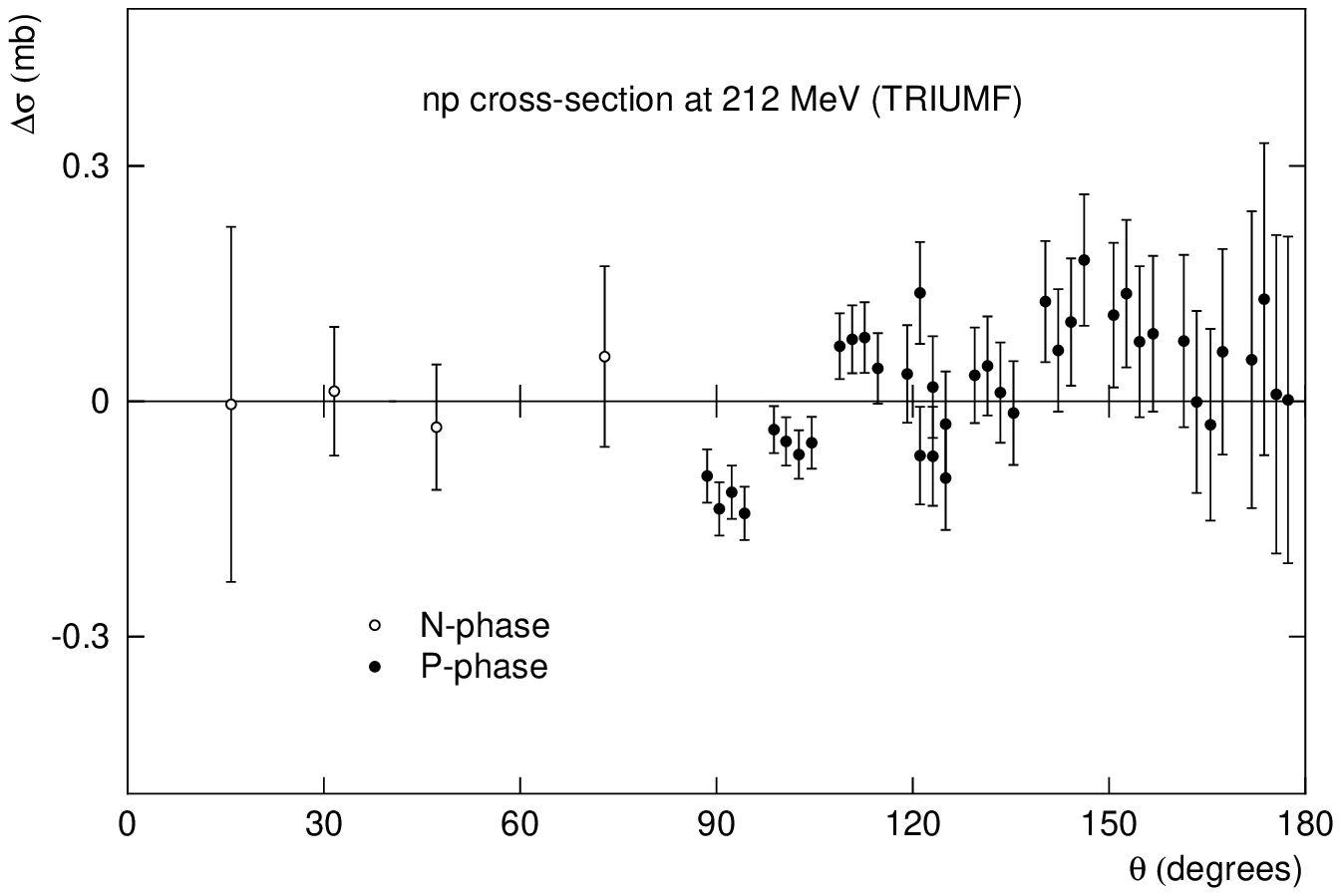}
\end{center}
\caption{The difference $\Delta\sigma(\theta)$ between the absolutely
         normalized TRIUMF data and PWA93 at 212 MeV.}
\label{fig:4}
\end{figure}

In the `N-phase' of the experiment the neutrons scattered in the
forward direction were detected by a carbon plate.
The absolute normalization was measured and
was said to have a 2\% uncertainty.
The differential cross section of the scattered neutrons
at 212 $\pm$ 3 MeV was measured at 4 angles in this N-phase.
In Fig.~\ref{fig:4} we show the difference between the
differential cross section, again normalized by us, and PWA93.
The forward data (N-phase) agrees very well with PWA93.
The fit is a little too good, indicating that the point-to-point
errors are probably not completely of statistical origin.
In PWA93 we obtained $\chi^2 = 0.5$
with $N_{abs} = 0.996(12)$ for these 4 data points.
At 319 $\pm$ 3 MeV 7 data points were measured in the N-phase.
In PWA93 these 7 points had $\chi^2 = 3.4$ with $N_{abs}= 0.917(12)$.

In the `P-phase' of the experiment the recoil protons were detected
either in a magnetic spectrometer or, at the lowest recoil energies,
in a thick scintillation counter.
At each setting of this spectrometer (or scintillation counter)
4 data points were measured covering
a proton angular region in the laboratory of about 4$^\circ$,
corresponding to a neutron angular region in the center-of-mass
system of about 8$^\circ$.
The absolute normalization in the P-phase was measured
with a 3.2\% uncertainty.

The differential cross section of the recoil protons at 212 MeV
was measured at 39 points in this P-phase; see Fig.~\ref{fig:4}.
This data consists of 9 subsets of 4 data points and 1 subset
of 3 points, because the subset at 122$^{\circ}$ had an additional
3 data points measured.
The differences between these 2 subsets, measured at different
times, is an indication of the accuracy with which the flux of the
incident neutron beam could be monitored.
It is striking that the statistical fluctuations within each subset
of 4 points are {\em smaller} than the point-to-point errors indicate.
It is clear from Fig.~\ref{fig:4} that the subsets at 91$^\circ$ and
101$^\circ$ can be made to fit excellently,
if we are allowed to change the normalization from subset to subset.

In PWA93 these 39 data points had $\chi^2=100$ with $N_{abs}=1.014(4)$,
while we expected $\langle\chi^2\rangle = 38 \pm 9$. Closer scrutiny
showed that the set of 4 data points around $90^\circ$ (neutron angles
88.57$^\circ$, 90.45$^\circ$, 92.34$^\circ$, and 94.29$^\circ$)
contributed 52 to this high value of $\chi^2$.
We therefore omitted these 4 points from the data set.
For the remaining 35 points we reached then the value $\chi^2$=44
with $N_{abs} = 1.008(4)$, which is quite satisfactory.
At 319 MeV the P-phase of the experiment measured 64 data points.
In PWA93 these 64 points were fitted with $\chi^2 = 78$ and
$N_{abs} = 1.007(3)$. Also this is a reasonable fit.

The measured absolute normalizations agree very well with
the absolute normalizations calculated by us, except for the N-phase
at 319 MeV, where the measured normalization appears to be 8.3\%, that
is 4 s.d., too high. Our calculated normalizations have a small error.
For the N-phase our error is 1.2\%, compared to 2\% for the measured
normalization, and for the P-phase our error is 0.4\%, which must be
compared to 3.2\% for the measured normalization.

\subsection{The Uppsala data}
At the neutron beam facility at the The Svedberg Laboratory (TSL)
in Uppsala the $np$ differential cross section was measured at two
energies: 96 MeV and 162 MeV.
In this paper, we will focus our attention on the 162 MeV
data~\cite{Eri95,Rah97,Rah98}.
The neutrons for this reaction were produced by the
$^7$Li($p,n$)$^7$Be reaction at 0$^\circ$.
This gave a quasi-monoenergetic neutron beam together
with an almost flat low-energy tail. The target consisted
of 5 CH$_2$ sheets and 2 carbon sheets interspaced by MWPC's.
The MWPC's were necessary to determine in which sheet the
reaction took place.
The carbon sheets were included to measure simultaneously
the protons produced on carbon, because the proton production from
the carbon nuclei in CH$_2$ needed to be subtracted.
The recoil protons were detected in a magnetic spectrometer
consisting of a dipole magnet and four drift chambers.
In each run the magnetic spectrometer had an angular acceptance
of about 15$^\circ$.
Five different settings of the spectrometer position were chosen
to cover the angular range of $\theta_{lab}= 0^\circ$ to
54$^\circ$ for the recoil proton.
This guaranteed a large angular overlap between the different settings.
The relative normalizations between these 5 different angular sets,
are {\em calculated} normalizations.
After the sets were relatively normalized, the data in the overlap
regions was averaged point by point.
In this way the relative differential cross section between 72$^\circ$
and 180$^\circ$ was obtained.
Next, this relative cross section needed to be absolutely normalized.
The integrated cross section $\sigma(a,b)$
in the interval from $a^{\circ}$ to $b^{\circ}$ is defined by
\begin{equation}
   \sigma(a,b) = 2\pi\int_a^b\sin\theta\,d\theta\,(d\sigma/d\Omega) \ .
\end{equation}
For the absolute normalization the Uppsala group used the average over
several PWA's of
\begin{equation}
   N_{abs}(upp) =
     \left[\frac{\sigma_{pwa}(72,180)}{\sigma_{un}(72,180)}\right]
     \left[\frac{\sigma_T(upp)}{\sigma_T(pwa)}\right] \ , \label{eq:upp}
\end{equation}
where $\sigma_{un}$ is the unnormalized integrated cross section
as determined in Uppsala and $\sigma_T$ is the total cross section at
162 MeV. For PWA93 the last fraction on the right-hand side
of Eq.~(\ref{eq:upp}) is 0.995. This indicates
clearly that the absolute normalization factor $N_{abs}$ is essentially
the first fraction on the right-hand side of Eq.~(\ref{eq:upp}).
It shows the importance of the PWA's in obtaining the calculated
absolute normalization of this data.

A few critical remarks can be made about this calculation.
First of all, in determining $\sigma_T(upp)$ at 162 MeV the Uppsala
group did not make use of all available relevant experimental data,
and, moreover, the data they did use, was not renormalized.
It is striking that the Uppsala group used an interpolated value for
the total cross section calculated by themselves, instead of using
a value determined in one of the standard PWA's.
Secondly, their way of normalizing does not take account of the
measured accuracy of each individual data point. The data points are
instead weighted by the factor $2\pi\sin\theta$.
A $\chi^2$-fit to the differential cross section over the
measured interval of a PWA would have been more appropriate.
Finally, the Uppsala group has made a big issue of the fact that their
differential cross section is {\em different} from the PWA's.
But then there is no good reason to assume that their integrated
cross section is the {\em same} as in the PWA's.
This conclusion casts doubts on their method of normalization.
Despite these critical remarks, we think that their calculated
absolute normalization is numerically not unreasonable.

\begin{figure}[t]
\begin{center}
\includegraphics{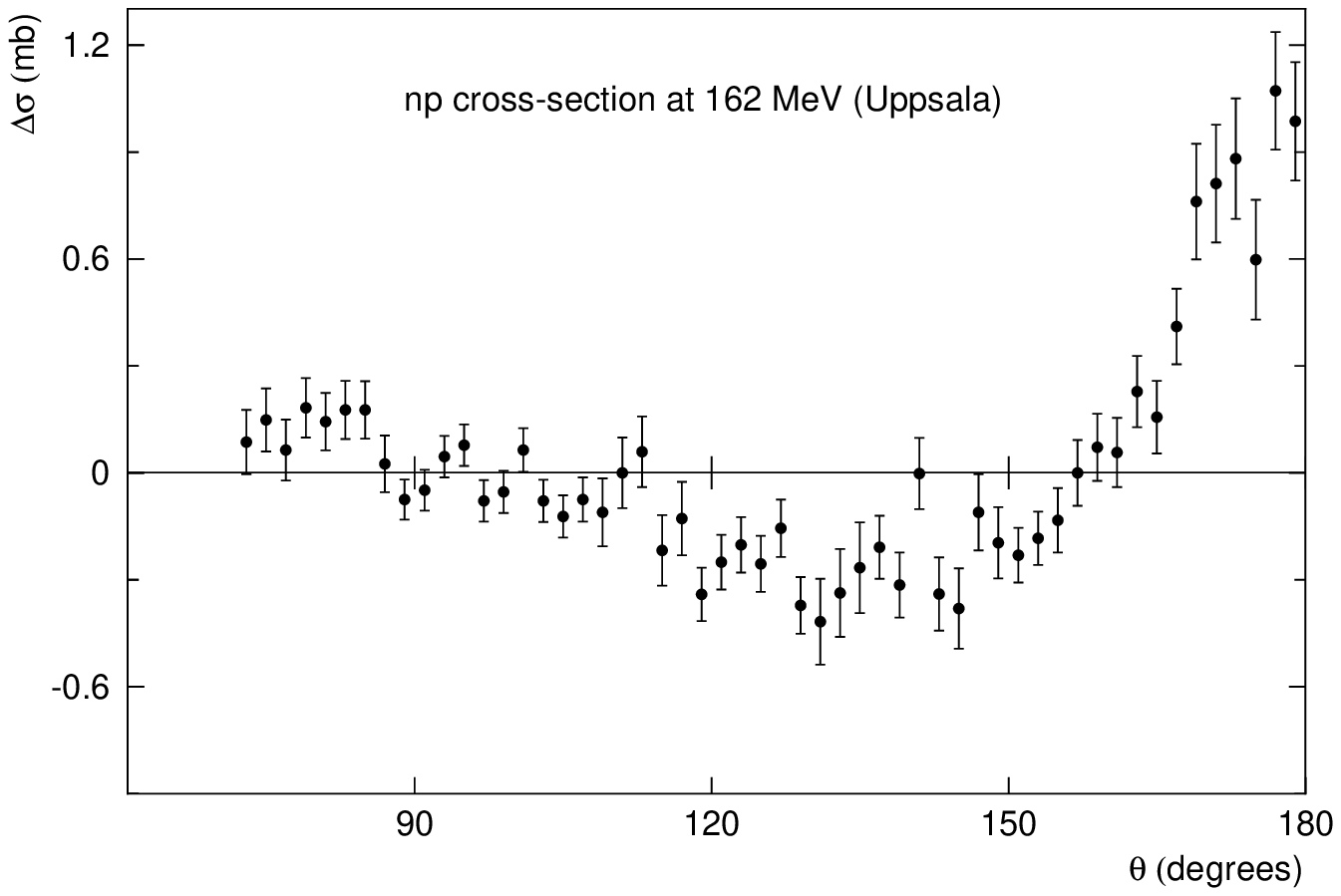}
\end{center}
\caption{The difference $\Delta\sigma(\theta)$ between the absolutely
         normalized Uppsala data and PWA93 at 162 MeV.}
\label{fig:5}
\end{figure}

In Fig.~\ref{fig:5} we show the difference between the
differential cross section, normalized by us, and PWA93.
Comparison with PWA93 gives for the 54 data $\chi^2 = 393$ and
$N_{abs} = 1.0109(24)$.
For 54 data one would expect $\langle\chi^2\rangle = 53 \pm 10$.
Our result is 33 s.d. away from this expectation value, which is
an {\em unbelievable} large discrepancy. This data set
will therefore not be included in our data base for PWA. The reason
is not that a large value for the $\chi^2$ for this data set will
give a larger value for the $\chi^2$/data point for the whole data base,
but such a large value is so unlikely, that our data base would not be a
Proper statistical ensemble anymore, when we would include this data set.

We have studied these 162 MeV Uppsala data in more detail.
The first time this was done~\cite{Ren98} with the originally published
data set~\cite{Eri95} of 31 data points between 119$^\circ$ and
180$^\circ$. In a PWA one expects for 31 points
the value $\langle\chi^2\rangle = 31 \pm 8$.
PWA93 with a data base of 2514 data gave $\chi^2$ = 292 for
these 31 data with the normalization $N_{abs}$ = 0.9820.
When we included the 31 data in the data base and did a refit the
$\chi^2$ on the 31 data dropped with 27.6, while at the same time
the $\chi^2$ on the rest of the data set increased by 7.4.
The difference with the expectation value amounts to 29 s.d. The
Virginia Tech PWA~\cite{Arn94} gives similar huge values for the $\chi^2$.
Such large values for the $\chi^2$ for only 31 data are statistically
so unlikely that something must be wrong.
The Uppsala group has always pointed to the LAMPF data as the culprit
and claims~\cite{Rah98} that this data dominates our solution and also
the SAID solution.
To study this claim we omitted all the LAMPF data~\cite{Bon78} from
our data base and included the 96 MeV and the 162 MeV Uppsala data.
Performing a PWA on this data set gave for the 31 data at 162 MeV
$\chi^2$ = 246. This amounts to a 27 s.d. error and is therefore
not much of an improvement.
It shows clearly that the claim~\cite{Rah98}, that our data base is
dominated by the LAMPF data, is incorrect.
Moreover, it shows that the Uppsala differential cross section is
in conflict not only with the LAMPF cross sections, but also with the
rest of the data base, including asymmetries, spin correlations, etc.

At that time, we had available in Nijmegen the first PWA~\cite{Klo94}
of the $np$ data alone, in which no input from $pp$ scattering was used.
This $np$-PWA94 was moreover different from PWA93, because it went up
to 500 MeV and therefore also inelasticities were included. For these
reasons, we used also this $np$-PWA94~\cite{Klo94} of
3964 $np$ data below 500 MeV to study the Uppsala data.
It gives for the 31 data the value $\chi^2$ = 293.
When we include the 31 data in the data base and refit,
the $\chi^2$ for these 31 points drops with 80,
while the $\chi^2$ on the rest increases by 34.7.
Finally, we dropped all 1482 differential cross sections with
$\theta > 119^{\circ}$ from the data base. Performing next a PWA on
this reduced data base gave $\chi^2$ = 174 for the 31 data points,
which corresponds to an 18 s.d. discrepancy. The isovector phases
are in this $np$-PWA94 determined by the $np$ data and {\em not} by
the more accurate $pp$ data. This allows for more freedom
in the isovector phases when fitting the $np$ data. As a consequence
a much larger drop in $\chi^2$ is possible than in the PWA93 case.
However, we still see that all $\chi^2$ values are very far away from
the expectation value $\langle\chi^2\rangle = 31 \pm 8$. This
proves once more that the 31 Uppsala data are in conflict,
not only with the LAMPF differential cross section data, but also
with the other data, including spin data, contained in the Nijmegen
and SAID data bases. The same conclusions can be drawn for the more
complete set of 54 data~\cite{Rah98}.

\begin{figure}[t]
\begin{center}
\includegraphics{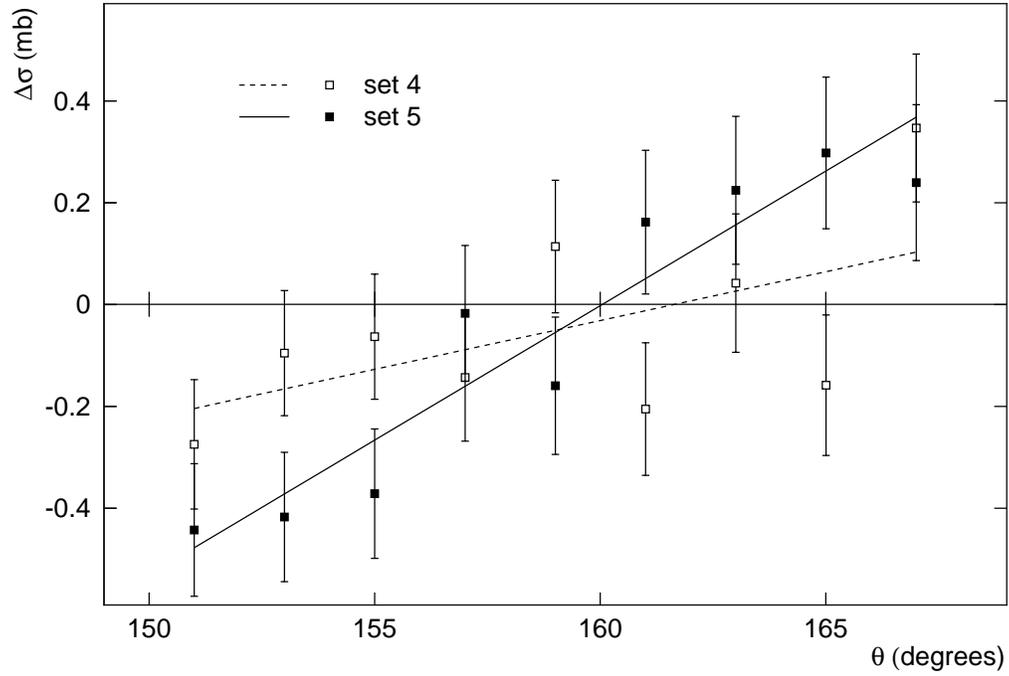}
\end{center}
\caption{The difference $\Delta\sigma(\theta)$ between the
         Uppsala data and PWA93 at 162 MeV in the overlap
         region between sets 4 and 5.}
\label{fig:6}
\end{figure}

A first hint why the Uppsala data produces such a large value of
$\chi^2$ can be obtained when we look at the data in the overlap
region between sets 4 and 5, see Fig.~\ref{fig:6}.
This overlap region runs from 151$^\circ$ to 167$^\circ$
and contains from each set 9 points.
In Fig.~\ref{fig:6} we plot the difference
$\Delta\sigma = \sigma(exp) - \sigma(pwa)$.
For each data set $i$ the function $\Delta\sigma$ is fitted
by a straight line $\Delta\sigma = a_i + b_i \: \theta$.
These fits are statistically quite acceptable.
For set 4 $\chi^2$ = 10.3 and for set 5 $\chi^2$ = 4.2.
In both cases 9 points were used in the fit.
For the slopes $b_i$ of the sets 4 and 5 we find
$b_4$ = 0.019(9) and $b_5$ = 0.053(9). We see that both slopes are
different from 0, therefore the slopes do not agree with PWA93.
Especially the slope of set 5 in this overlap region is almost 6
s.d. away from the slope predicted by PWA93.
We believe that this is too large a difference to be acceptable.

>From Fig.~\ref{fig:6} we see that these straight lines cross each other
at 159$^\circ$, which is the middle of the overlap region.
This is in agreement with the fact that the Uppsala group took the
relative normalization between the sets such, that in the overlap region
the differential cross section is as much as possible continuous.
They could not require that the slope was also continuous.
The slopes $b_4$ and $b_5$ of set 4 and 5 in the overlap region turn
out to be rather different.
We find $b_5 - b_4$ = 0.034(13). This is a surprising result, since
the difference should be consistent with zero. We think that 2.6 s.d.
is too large a difference to be called `consistent with zero.'
Therefore, in the overlap region the sets 4 and 5 appear to be in
disagreement with each other~\cite{Error}.

\section{Relative normalizations} \label{REL}
In this Section we will look in more detail at the way the relative
normalizations in the various data sets are performed, and we will
point out the way we think these relative normalizations could be done.
It is remarkable that for the various data sets under consideration
in this paper each one requires a different treatment of the relative
normalization.

\subsection{The LAMPF data revisited}

In the LAMPF experiment the whole angular region from proton laboratory
angle 0$^\circ$ to about 30$^\circ$ was covered by 7 different
settings of the central angle of the spectrometer.
Simultaneously the incident neutron flux on the liquid-hydrogen target
was measured. This allowed the LAMPF group to determine experimentally
the relative normalizations of these 7 angular sets.
For the lower energies ($T_{lab} < 275$ MeV) the angular sets belonging
to different central settings can be clearly identified. For the higher
energies ($T_{lab}>275$ MeV) these angular sets started to overlap. At
present, it is not possible anymore to disentangle these
overlaps~\cite{Bon00}.
Therefore, we will divide the LAMPF set of data at 11 energies with
$T_{lab}<350$ MeV into two groups.
The group LAMPF-I contains the data at the
7 energies from $T_{lab}$ = 162 MeV to $T_{lab}$ = 266 MeV for which
the angular sets can still clearly be identified, while the group
LAMPF-II contains the data at those energies for which the central
angular settings cannot uniquely be recovered anymore.

The group LAMPF-II contains 312 data and is fitted
with $\chi^2$ = 327. In PWA93 the 4 absolute normalizations
$N_{abs} \pm dN_{abs}$ at the 4 different energies
from $T_{lab}$ = 284.8 MeV to $T_{lab}$ = 344.3 MeV were determined.
The results are displayed in Table~\ref{tab:2}.
The expectation value is $\langle\chi^2\rangle=308\pm25$.
This shows that in PWA93 the fit to the group LAMPF-II is quite good.
We emphasize again the high accuracy $dN_{abs}$ with which these
{\em calculated} absolute normalizations could be determined in PWA93.

The group LAMPF-I contains 337 data.
As pointed out before, in PWA93 the data set at 194 MeV with 42
data points was removed from the data base, because it had
a too high $\chi^2$. We were therefore left with 295 data.
This gives the expectation value $\langle\chi^2\rangle=289\pm24$.
In PWA93 we obtained $\chi^2$ = 303.
When we now restored these 42 data points back into the data base,
and analyzed the group LAMPF-I, we found that the datum at 162 MeV and
the neutron angle of 174.32$^\circ$ was more than 3 s.d. off. Therefore,
we removed this point from our data set and redid the analysis,
now with 336 data points.
When we fit only the 7 absolute normalizations $N_{abs}$, we expect
$\langle\chi^2\rangle=329\pm26$ and we obtain $\chi^2$ = 373.

Next, we introduced relative normalizations $N_{rel}(l)$,
with $l$ going from 1 to 7,
for each of the 7 central angular settings of the magnetic spectrometer.
Here $l$ = 1 corresponds to the largest proton angle and therefore the
smallest neutron angle, while $l$ = 7 is the most backward direction.
We normalize $N_{rel}(4)$ = 1.
In the experiment these relative normalizations were measured with
an error $dN_{rel}(l)$. The exact value of $dN_{rel}(l)$ is at present
unknown, but probably it was of the order of 1\% or less~\cite{Bon00}.
In our analysis we tried 4 values: 0\%, 0.5\%, 1.0\%, and floated.
The value of 0\% corresponds to PWA93, where we did not try to
study these relative normalizations, so we took exactly the values as
given by the experimentalists, with no errors. Thus, the relative
normalization in that case was 1.0 $\pm$ 0.0.
Next to these relative normalizations of the different central angular
settings there are of course the absolute normalizations $N_{abs}(k)$,
with $k$ going from 1 to 7. Here $k$ = 1 corresponds to the lowest energy
162.0 MeV.
The contribution to $\chi^2(data)$ of the differential cross sections
$\sigma_{kl}(\theta_i,exp)$ with their errors $d\sigma_{kl}(\theta_i,exp)$
is given by
\begin{equation}
    \chi^2(data) = \sum_{k,l,i}
    \left[\frac{N_{abs}(k)\:N_{rel}(l)\:\sigma_{kl}(\theta_i,exp) -
                                        \sigma_{kl}(\theta_i,pwa)}
    {N_{abs}(k)\:N_{rel}(l)\:d\sigma_{kl}(\theta_i,exp)} \right]^2 \ ,
\end{equation}
while the contribution to $\chi^2(norm)$ of the absolute and relative
normalizations is
\begin{equation}
    \chi^2(norm) =
       \sum_{k}\left[\frac{N_{abs}(k)-1}{dN_{abs}(k)}\right]^2 +
       \sum_{l}\left[\frac{N_{rel}(l)-1}{dN_{rel}(l)}\right]^2 \ .
\end{equation}

In Table~\ref{tab:3} the $\chi^2$-values for the different cases are
presented together with the number of data per energy and the number
of degrees of freedom.
In Table~\ref{tab:4} we present the relative and absolute normalizations.
Our conclusion from Table~\ref{tab:3} is that the case of 1\% relative
normalization error turns out to be quite good.
It is an improvement over PWA93 of 30 in the value of $\chi^2(total)$,
and of 36 in the value of $\chi^2(data)$.
What is more important is what happens to the 194.5 MeV data.
For these 42 data points the 3 s.d. rule requires that
         $ 20 < \langle \chi^2 \rangle < 71 $,
In the original PWA93 we obtained for this data $\chi^2 = 76$.
This meant that this set had to be removed from the PWA93 data base.
In the present analysis we obtain $\chi^2 = 64$, which is significantly
below the 3 s.d. limit and therefore we can keep this set.
For the 162 MeV data we see that the $\chi^2$ drops from 63 for 43 points
in PWA93, to 56 for 42 points in the same PWA93, and to 48 for these
42 points when we take the relative normalizations with 1\% error into
account. This shows that this 162.0 MeV data appears to be quite good.
For the 43 data at 211.5 MeV the drop in $\chi^2$ is from 31 to 27
when we take the relative normalizations with 1\% error into account.
Thus also this 211.5 MeV data appears to be quite good.

In PWA93 we included 607 of the 650 LAMPF data and obtained
$\chi^2(data) = 630$ with $F=\chi^2(data)/N(data)=1.038$.
In the present analysis we included 648 of this LAMPF data and
obtained $\chi^2(data)=664$ with $F=\chi^2(data)/N(data) = 1.025$.
The improvement in quality of the present analysis over PWA93,
due to the introduction of the relative normalizations, is
seen in the fewer data omitted from the data set (2 versus 43)
and in the better $F$-ratio (1.025 versus 1.038).

\subsection{The TRIUMF data revisited}
In PWA93 we discarded the 4 most forward points (neutron angles)
of the P-phase of the experiment at 212 MeV. These correspond to the most
backward proton angles and therefore to the lowest energies of the recoil
protons. These angles are the hardest ones to measure, and it is therefore
understandable that these points suffer from the largest systematic errors.
Before we study the relative normalizations, we will reinstate these 4
most forward points to the data set.
This data set contains then 10 subsets.
When we assume the error in the relative normalizations to be very
large, i.e. we float these relative normalizations, we obtain
$\chi^2 = 10.2$ on this data. The expected value is
$\langle \chi^2 \rangle = 29$ and the lower 3 s.d. limit is 12.2.
We see that the $\chi^2$ has become too low, hence in this case
this data should be discarded.
The reason for this is that the point-by-point errors are not
purely statistical, but contain systematic components.

When we assume a relative normalization error of 0.5\% and an absolute
normalization error of 0.2\%, we obtain
$\chi^2(data) = 53.9$ and $\chi^2(norm) = 13.1$.
However, looking in more detail at the results we see that the group of
4 most forward data contributes 26.9 to $\chi^2(data)$.
Therefore, we should float the relative normalization of only this subset.
Then we obtain $\chi^2(data) = 29.6$ and $\chi^2(norm) = 7.7$.
The most forward subset of 4 data contributes now only 1.0 to this value
of $\chi^2(data)$. The 3 s.d. lower limit for the expectation value of
$\chi^2$ is in this case 18.9.
The data is therefore statistically acceptable.

What have we gained by taking account of the relative normalizations?
First of all, we do not have to discard the most forward subset when we
allow the relative normalization of this subset to float.
Thus, instead of 35 data in PWA93 we can now use all 39 data of the P-phase.
Secondly, we reach now the value $\chi^2(data) = 29.6$ for these 39 points,
while in PWA93 we reached $\chi^2(data) = 44$ for 35 points.

\subsection{The Uppsala data revisited}
\begin{figure}[t]
\begin{center}
\includegraphics{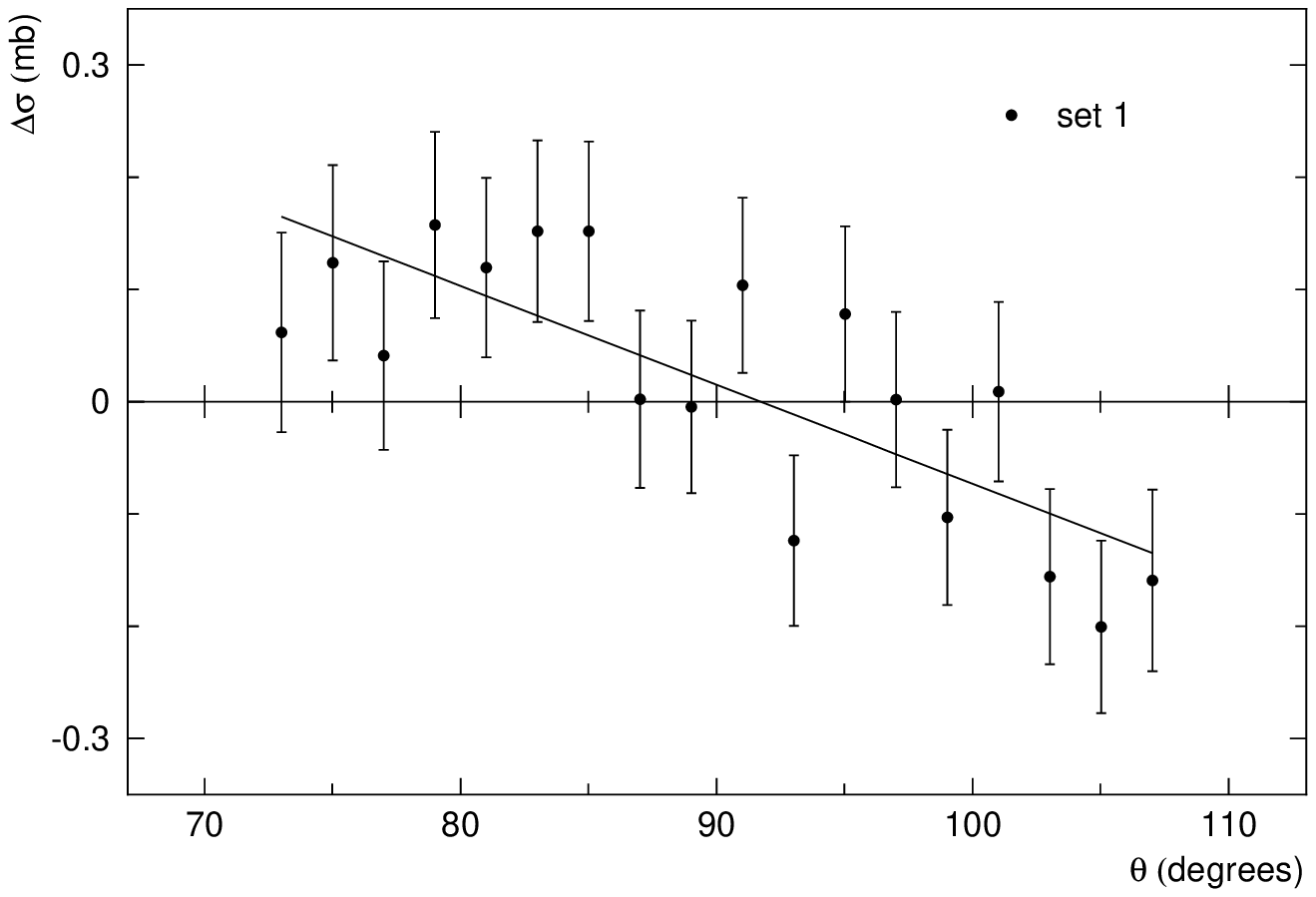}
\end{center}
\caption{The difference $\Delta\sigma(\theta)$ between the absolutely
         normalized set 1 of the Uppsala data and PWA93 at 162 MeV.}
\label{fig:7}
\end{figure}

\begin{figure}[t]
\begin{center}
\includegraphics{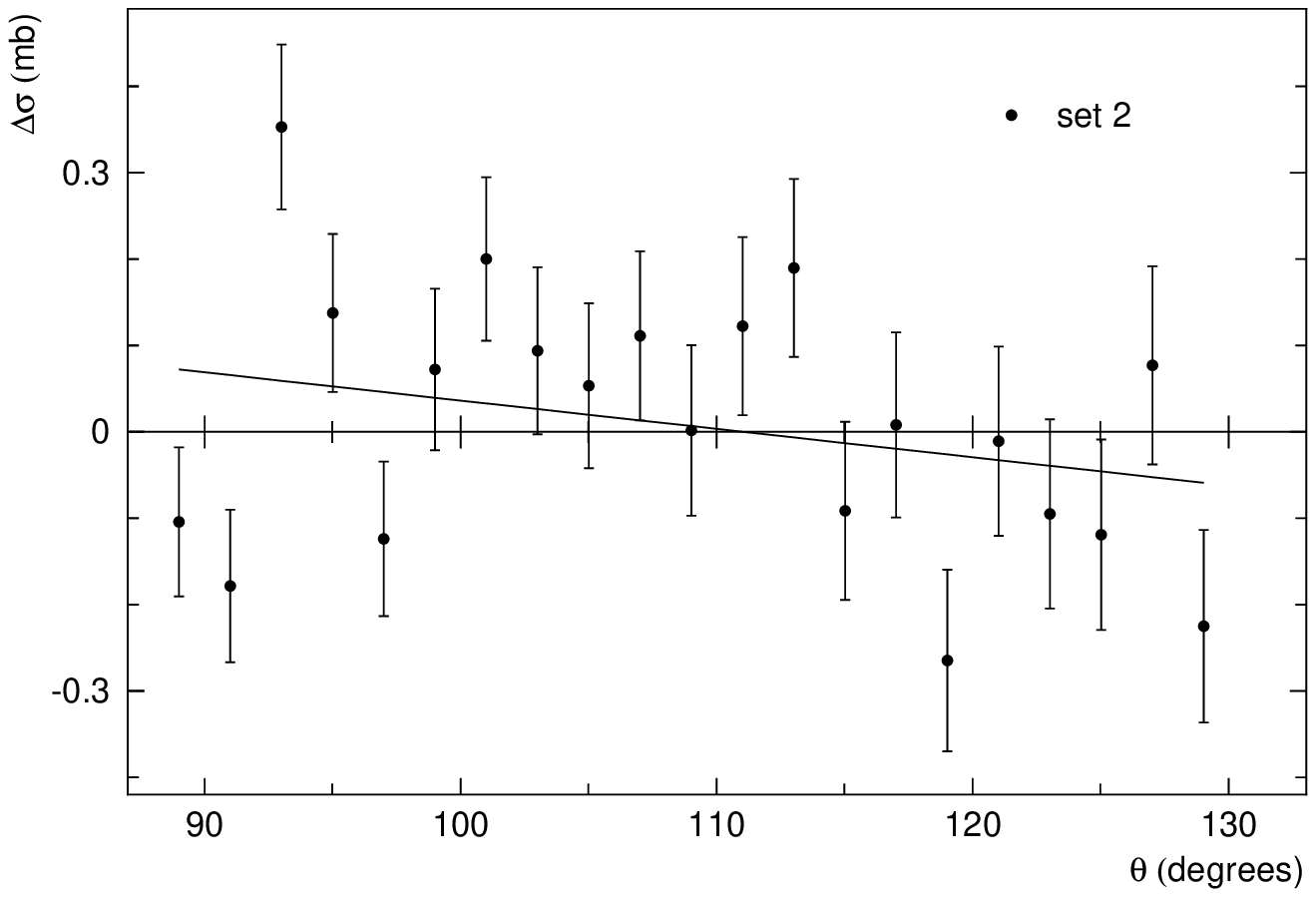}
\end{center}
\caption{The difference $\Delta\sigma(\theta)$ between the absolutely
         normalized set 2 of the Uppsala data and PWA93 at 162 MeV.}
\label{fig:8}
\end{figure}

\begin{figure}[t]
\begin{center}
\includegraphics{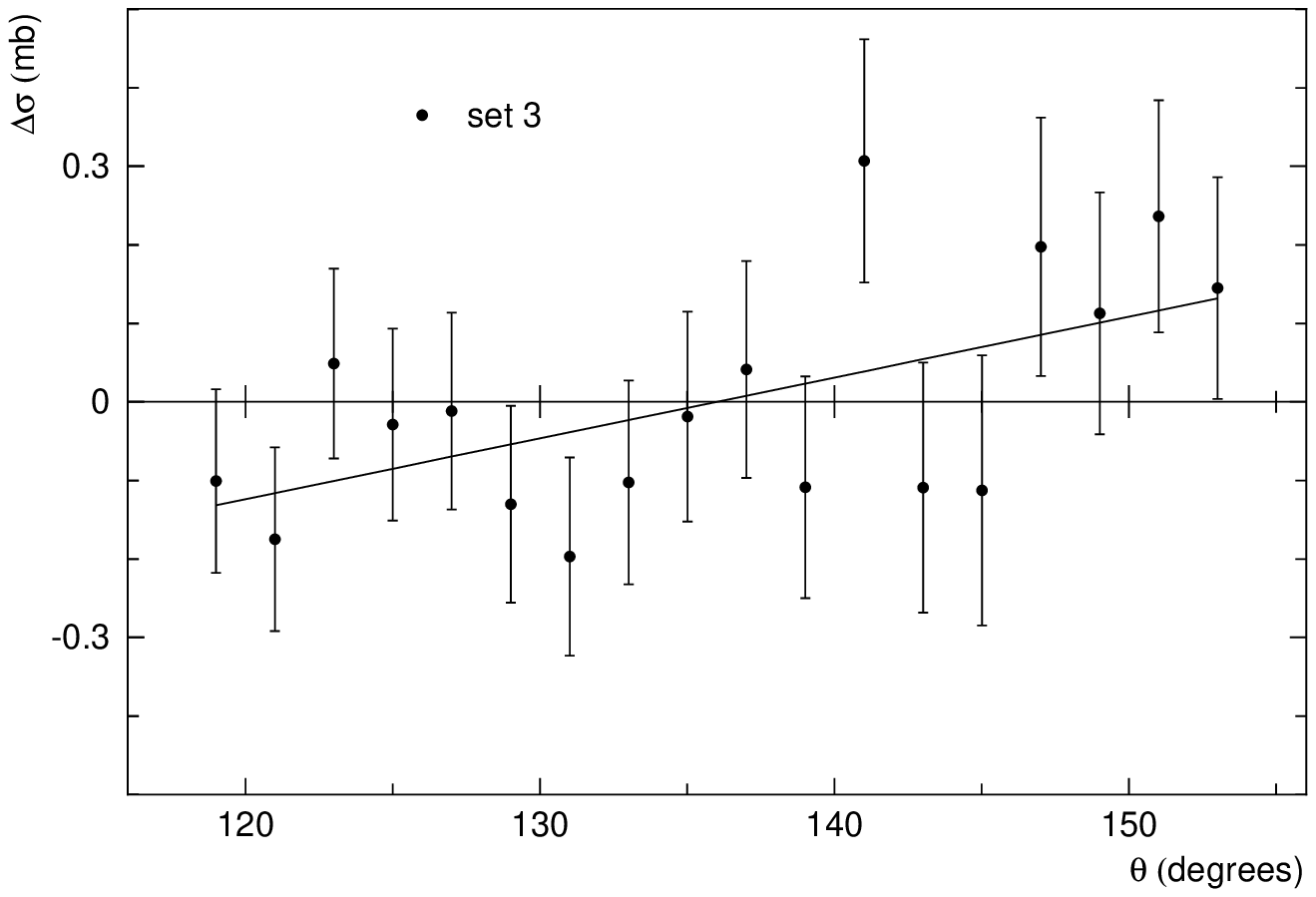}
\end{center}
\caption{The difference $\Delta\sigma(\theta)$ between the absolutely
         normalized set 3 of the Uppsala data and PWA93 at 162 MeV.}
\label{fig:9}
\end{figure}

\begin{figure}[t]
\begin{center}
\includegraphics{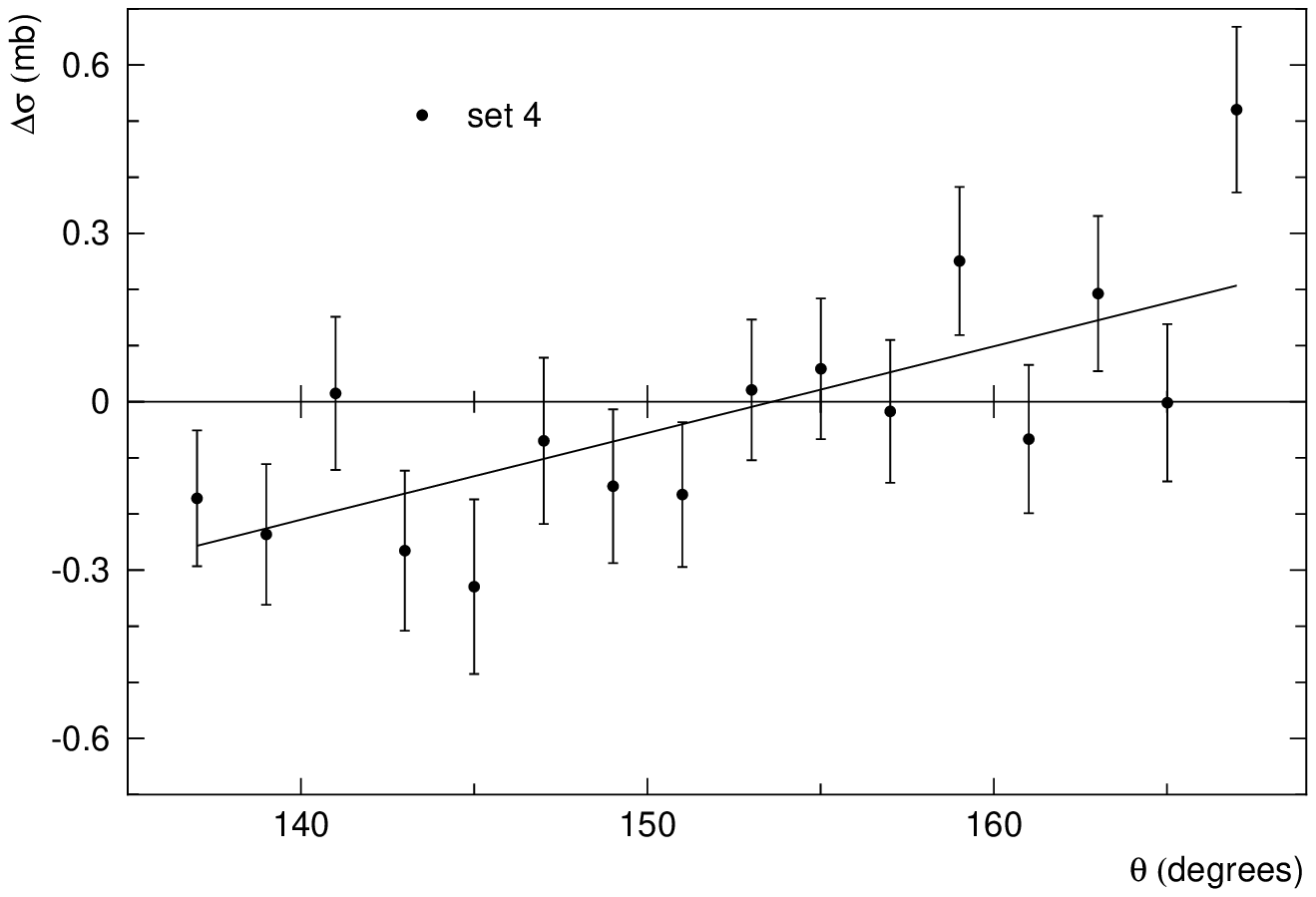}
\end{center}
\caption{The difference $\Delta\sigma(\theta)$ between the absolutely
         normalized set 4 of the Uppsala data and PWA93 at 162 MeV.}
\label{fig:10}
\end{figure}

\begin{figure}[t]
\begin{center}
\includegraphics{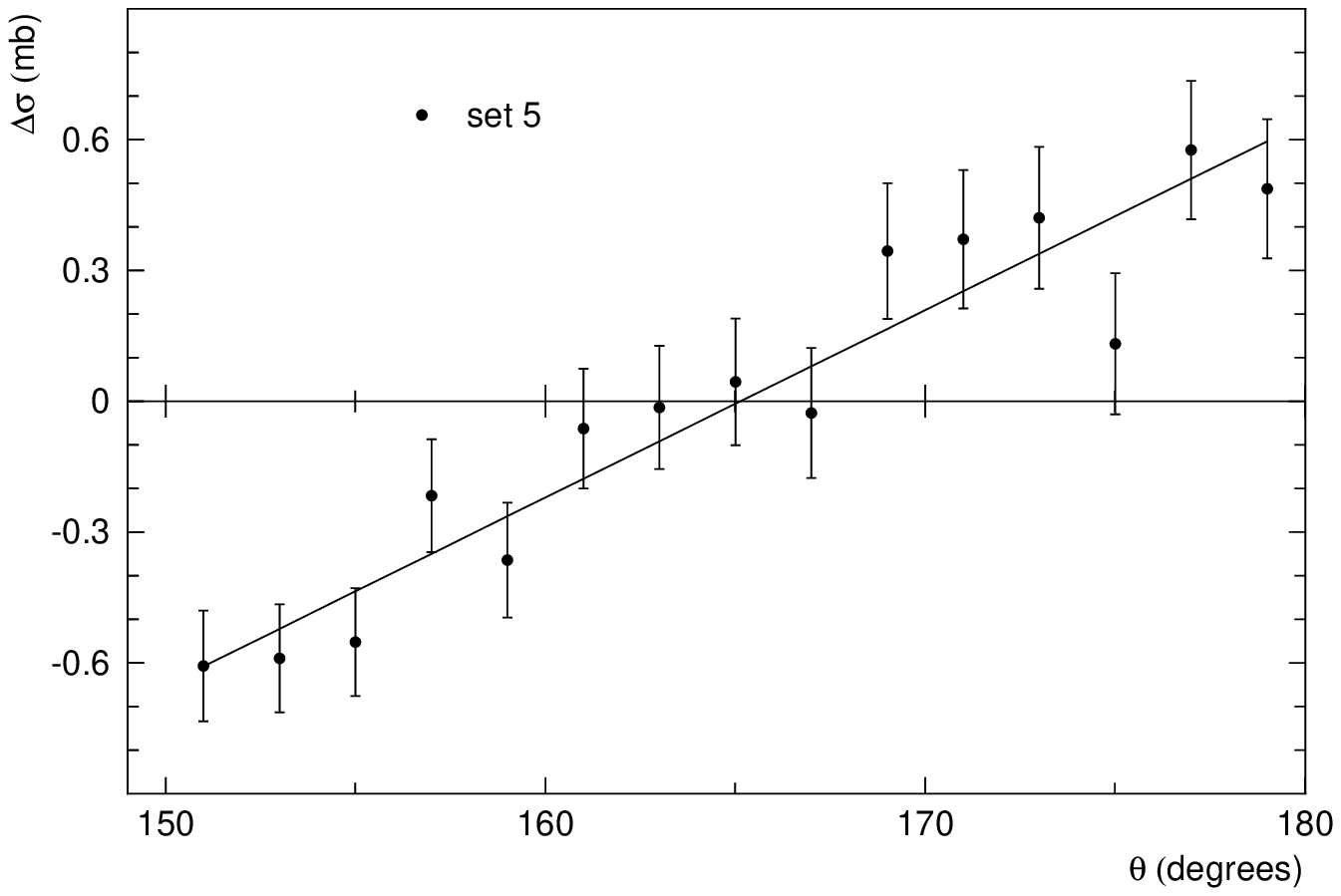}
\end{center}
\caption{The difference $\Delta\sigma(\theta)$ between the absolutely
         normalized set 5 of the Uppsala data and PWA93 at 162 MeV.}
\label{fig:11}
\end{figure}

Let us start with studying the individual Uppsala data sets $i$,
where $i = 1,\dots,5$, each covering a different angular region.
In Fig.~\ref{fig:7} to Fig.~\ref{fig:11} we plot for each set the
difference $\Delta\sigma(i) = N_{abs}(i)\sigma(exp,i)-\sigma(pwa,i)$,
where we used PWA93. For each
individual data set the value of $\chi^2$ is given in Table~\ref{tab:5},
together with the calculated absolute normalization $N_{abs}$ and the
number of data points $N(data)$.
The expectation value for $\chi^2$ of these sets is approximately
$\langle\chi^2\rangle = (N(data)-1)\pm\sqrt{2N(data)}$.
For the 3 s.d. upper bound we actually do not use this formula,
but we use an exacter formula, which produces a slightly larger value.
The $\chi^2$-values of the sets 2, 4, and 5 are more than 3 s.d. away
from their expectation values, while set 1 is almost
3 s.d. away and only set 3 has an acceptable $\chi^2$-value.
Using the 3 s.d. rule we must certainly omit 3 of the 5 sets from the
data base. Because the $\chi^2$ of one of the remaining two sets is also
large, it is perhaps advisable to disregard the complete Uppsala data set.

To understand better what is going on, we fitted for each individual
data set a straight line
$\Delta\sigma(i) = a(i)+b(i)\:\theta$ through
the $N(data)$ points of the set. The values for $a$, $b$, and
$\chi^2(sl)$ for the straight-line fits are also given in Table~\ref{tab:5}.
There is a dramatic improvement in $\chi^2(sl)$ over $\chi^2$ for the
sets 1, 4, and 5. This indicates very large {\em systematic} deviations
from PWA93 in these 3 data sets.
The slopes of these 3 individual data sets disagree strongly
with the slopes predicted by PWA93.
The straight-line fit to set 2 did not show a marked
improvement in the value of $\chi^2(sl)$. The unacceptably high value
for $\chi^2(sl)$ in set 2 indicates that this set should certainly be
removed from our data set.
The straight-line approximation gives also a definite improvement in
the $\chi^2$-value of set 3.
This indicates that also this set is perhaps infected with
systematic effects.
Looking at all this we come to the conclusion, that it is perhaps best
to disregard {\em all} the Uppsala data.

\begin{figure}[t]
\begin{center}
\includegraphics{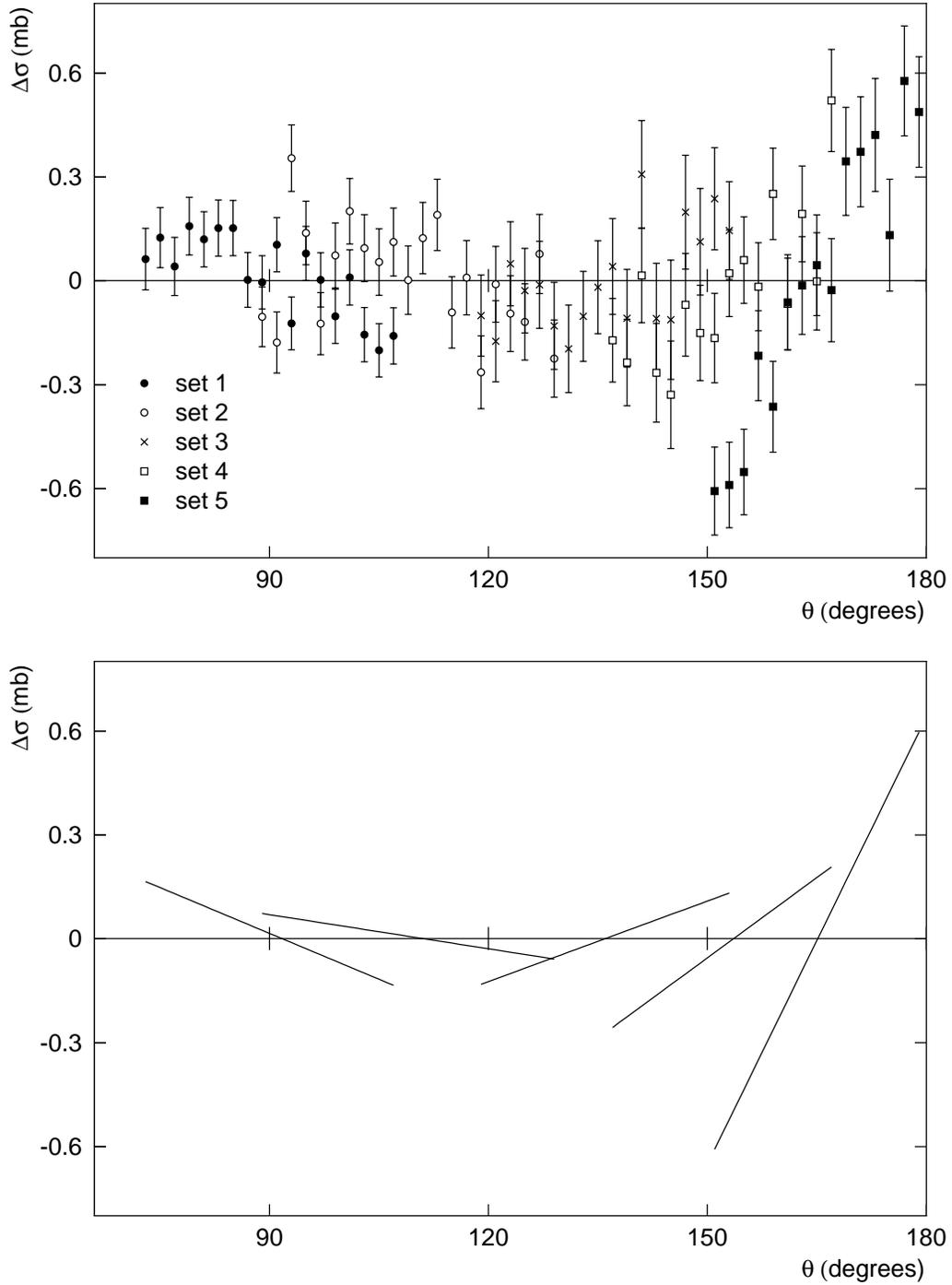}
\end{center}
\caption{The difference $\Delta\sigma(\theta)$ between the
         Uppsala data (each set separately absolutely normalized)
         and PWA93 at 162 MeV. In the bottom panel the data sets are
         replaced by their approximation by straight lines.}
\label{fig:12}
\end{figure}

Let us now come to the relative normalization of these data sets
in the case that the data is not discarded. In our PWA,
we would determine for each individual data set separably an
absolute normalization. We have then 5 separate calculated absolute
normalizations and no relative normalizations.
The result is plotted in Fig.~\ref{fig:12}. We have also plotted the
straight-line approximations to the sets.
>From Table~\ref{tab:5} we see that for the complete data set we have
88 points with $\chi^2 = 257$ with respect to PWA93.
For the expectation value of $\chi^2$ we find
$\langle\chi^2\rangle=83\pm13$. We could reduce this improbably high
value of $\chi^2$ by omitting individual data points with too high
(more than 3 s.d.) values of $\chi^2$, that is, omitting so-called
{\em outliers}. This is an acceptable procedure when these outliers
are isolated events. caused by unexplained errors in the measurement.
In this case, however, it is not acceptable, because these outliers are
caused by systematic effects.

\begin{figure}[t]
\begin{center}
\includegraphics{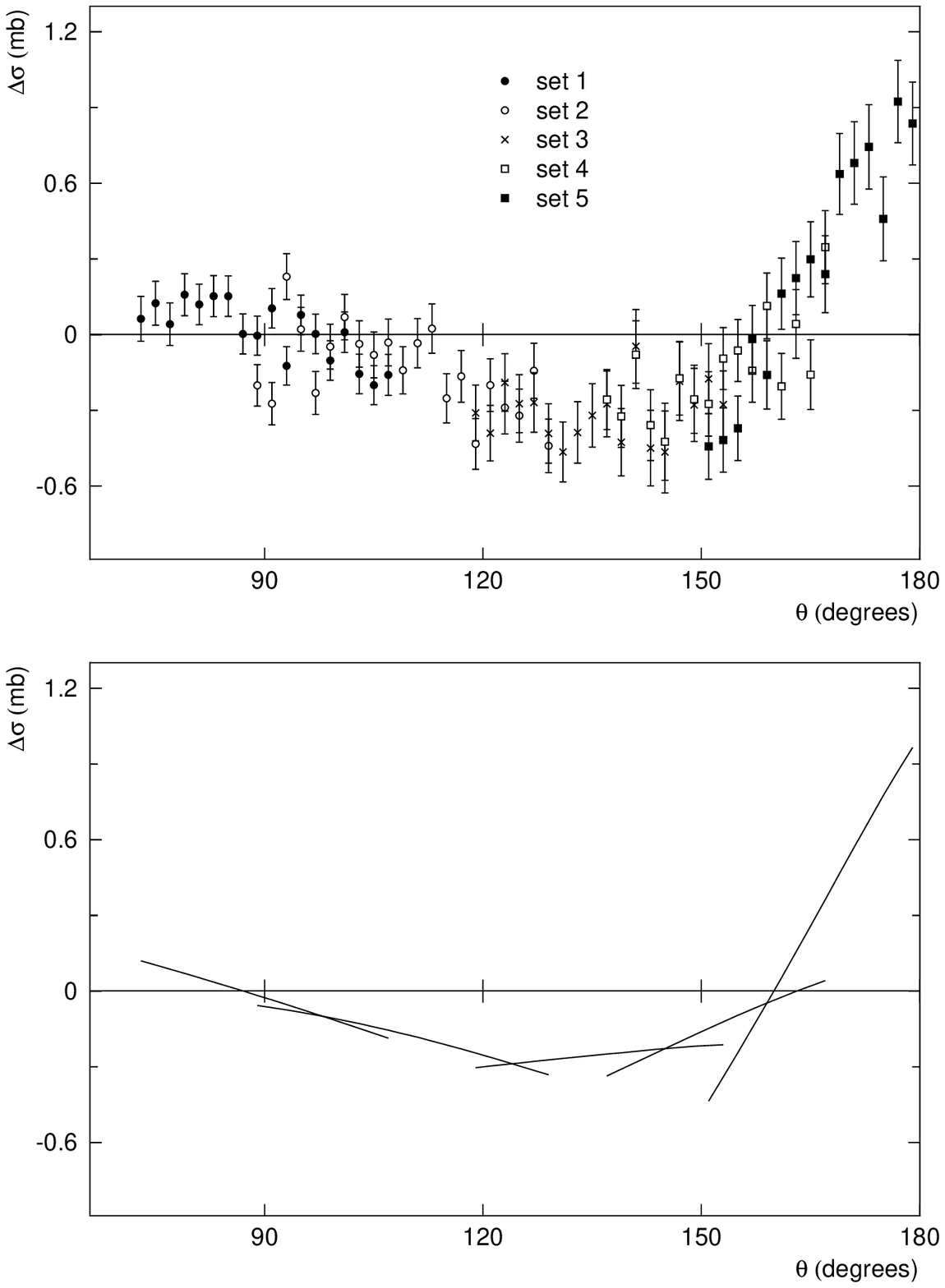}
\end{center}
\caption{The difference $\Delta\sigma(\theta)$ between the Uppsala data
         (each set relatively normalized as done by the Uppsala group)
         and PWA93 at 162 MeV. In the bottom panel the data sets are
         replaced by their approximation by straight lines.}
\label{fig:13}
\end{figure}

Finally, let us look at the way the relative normalizations were
determined by the Uppsala group. We demonstrate this with the
straight-line approximations to the individual data sets.
We require these straight lines to intersect in the middle of each
overlap region. In this way a more-or-less continuous, unnormalized
differential cross section is obtained.
This unnormalized cross section needs to be absolutely normalized.
The result is plotted in Fig.~\ref{fig:13}.
For the original set of 88 points as given by the Uppsala group and
as plotted in Fig.~\ref{fig:13} we find that $\chi^2 = 470$.

In this Subsection we have shown that the Uppsala data contains unexplained
large systematic deviations from PWA93. The manner in which the Uppsala
group performed the relative normalization of their data enhanced the
negative effects of these systematic deviations.
We found that the value of $\chi^2$ for these Uppsala data
can be reduced from the unacceptably high value of 470 to
the much better, but still unacceptably high, value of 257 by just
applying another way of relative normalization of this data.
Because of the large {\em systematic} deviations from PWA93
we cannot prune the data by omitting outliers. This means that, in
this way, we cannot reduce the value of $\chi^2$ any further and that
therefore we must omit the Uppsala data set from the $np$ data base.

\section{Conclusions} \label{CON}
We have shown that a more careful study of the relative normalization
of $np$ differential cross sections leads to a significantly improved
treatment of this data. The improvement shows up in several ways.
First, less data needs to be discarded in this analysis compared to
PWA93 in order to comply with the 3 s.d. rule.
For the LAMPF data we need to omit only 2 data points instead of the 43
data points in PWA93. From the TRIUMF data at 212 MeV we could keep all
data now, while in the PWA93 one needed to omit 4 data points. We were,
in this manner, unable to improve the fit to the Uppsala data so much,
that this data becomes acceptable.
Secondly, the introduction of relative normalizations, which are fitted
to the data, leads to a definite improvement in the values of $\chi^2$.
For the LAMPF data below 350 MeV we find an improvement in
$F=\chi^2(data)/N(data)$ from 1.038 to 1.025,
for the P-phase of the TRIUMF data at 212 MeV we see an improvement in
$F$ from 1.26 to 0.76, and for the 88 Uppsala data at 162 MeV we see an
improvement in $F$ from 5.34 to 2.92.
We would like to point once more to the accuracy with which the relevant
normalizations can be determined. Depending on the data set this runs from
0.4\% to 0.8\%. We would like to recommend to the experimentalists, that
when it is too difficult and/or too expensive to determine the absolute
normalization for their data, they should consider measuring the
unnormalized differential cross section anyway and leave the normalization
to the PWA's. This could save them a lot of headaches.

\acknowledgements
First of all we thank Dr.R.A.M. Klomp for his contributions in the very
beginning of this work.
We thank B.E. Bonner for clarifying communications about the LAMPF data,
D. Axen and D.F. Measday for helpful discussions regarding the TRIUMF data,
N. Olsson for sending us the Uppsala data, and Th.A. Rijken for helpful
discussions.
Finally, we thank J. Blomgren for his enlightening e-mail.
The research of R.T. was made possible by a fellowship of the
Royal Netherlands Academy of Arts and Sciences (KNAW).

\begin{table}
\begin{tabular}{ccccc}
$T_{lab}$ (MeV)  & $N_{abs}(dN_{abs})$   & $\chi^2$ & $N_{data}$ & removed \\
\hline    162.0  &  1.092(7)      &  63   &  43  & \\
          177.9  &  1.083(7)      &  47   &  44  & \\
          194.5  &  1.078(7)      &  76   &  42  & all \\
          211.5  &  1.063(7)      &  31   &  43  & \\
          229.1  &  1.058(7)      &  65   &  49  & \\
          247.2  &  1.042(7)      &  39   &  53  & \\
          265.8  &  1.029(6)      &  59   &  63  & \\
          284.8  &  1.053(5)      &  80   &  73  & \\
          304.2  &  1.003(4)      &  80   &  79  & \\
          324.1  &  1.057(5)      &  92   &  81  & \\
          344.3  &  1.036(5)      &  84   &  80  & 131.51$^\circ$ \\ \hline
          total  &                & 716   & 650  & \\
\end{tabular}
\caption{The calculated absolute normalization $N_{abs}$ and
         $\chi^2$ for the LAMPF data.}
\label{tab:1}
\end{table}

\begin{table}
\begin{tabular}{cccc}
$T_{lab}$ (MeV) &  $N_{abs}(dN_{abs})$   & $\chi^2$ & $N_{data}$  \\
\hline
        284.8  &        1.053(5)        &    80    &    73       \\
        304.2  &        1.003(4)        &    80    &    79       \\
        324.1  &        1.057(5)        &    92    &    81       \\
        344.3  &        1.035(5)        &    75    &    79       \\
\hline
        total  &                        &   327    &   312       \\
\end{tabular}
\caption{The calculated absolute normalization $N_{abs}$ and $\chi^2$
         for the LAMPF-II data.}
\label{tab:2}
\end{table}

\begin{table}
\begin{tabular}{cccccc}
   $T_{lab}$ (MeV) &  0\%  & 0.5\% & 1\%   & floated & $N_{data}$ \\
\hline     162.0   &   56  &   50  &   48  &   47    &   42       \\
           177.9   &   47  &   41  &   39  &   38    &   44       \\
           194.5   &   76  &   68  &   64  &   61    &   42       \\
           211.5   &   31  &   28  &   27  &   27    &   43       \\
           229.1   &   65  &   62  &   62  &   62    &   49       \\
           247.2   &   39  &   39  &   40  &   42    &   53       \\
           265.8   &   59  &   56  &   57  &   59    &   63       \\ \hline
  $\chi^2(data)$   &  373  &  344  &  337  &  336    &            \\
  $\chi^2(norm)$   &   $-$ &    9  &    6  &    0    &            \\
  $\chi^2(total)$  &  373  &  353  &  343  &  336    &            \\
  $N(dof)$         &  329  &  329  &  329  &  323    &            \\
\end{tabular}
\caption{The $\chi^2$-values for the 336 LAMPF-I data for
         various relative normalization errors.}
\label{tab:3}
\end{table}

\begin{table}
\begin{tabular}{ccccc}
                 &  0\%      & 0.5\%     & 1\%       &  floated   \\ \hline
   $N_{rel}(1)$  & 1.0       & 1.005(4)  & 1.011(7)  &  1.018(8)  \\
   $N_{rel}(2)$  & 1.0       & 1.003(4)  & 1.005(6)  &  1.007(6)  \\
   $N_{rel}(3)$  & 1.0       & 0.997(4)  & 0.996(6)  &  0.996(5)  \\
   $N_{rel}(4)$  & 1.0       & 1.000(4)  & 1.000(7)  &  1.000(7)  \\
   $N_{rel}(5)$  & 1.0       & 0.999(4)  & 0.998(6)  &  0.997(6)  \\
   $N_{rel}(6)$  & 1.0       & 0.992(4)  & 0.987(6)  &  0.984(5)  \\
   $N_{rel}(7)$  & 1.0       & 1.011(3)  & 1.015(5)  &  1.018(4)  \\ \hline
   $N_{abs}(1)$  & 1.089(6)  & 1.085(6)  & 1.083(6)  &  1.081(6)  \\
   $N_{abs}(2)$  & 1.083(7)  & 1.080(7)  & 1.078(6)  &  1.077(6)  \\
   $N_{abs}(3)$  & 1.078(7)  & 1.075(7)  & 1.073(6)  &  1.071(6)  \\
   $N_{abs}(4)$  & 1.063(7)  & 1.061(6)  & 1.059(6)  &  1.058(6)  \\
   $N_{abs}(5)$  & 1.058(7)  & 1.055(6)  & 1.053(5)  &  1.052(5)  \\
   $N_{abs}(6)$  & 1.042(7)  & 1.039(6)  & 1.038(5)  &  1.036(5)  \\
   $N_{abs}(7)$  & 1.029(6)  & 1.026(6)  & 1.024(5)  &  1.023(5)  \\
\end{tabular}
\caption{The calculated relative and absolute normalizations $N_{rel}(l)$
         and $N_{abs}(k)$ and their errors for the LAMPF-I data.}
\label{tab:4}
\end{table}

\begin{table}
\begin{tabular}{ccrcccc}
   set & $N$(data) & $\chi^2$ & $\chi^2(sl)$ & $N_{abs}$  & $a$ & $b$    \\
\hline
    1  &  18  &   38  &  15  & 1.000(8) &  0.81(17) &$-$0.0088(19) \\
    2  &  21  &   49  &  46  & 1.051(7) &  0.37(19) &$-$0.0033(18) \\
    3  &  18  &   18  &  12  & 1.062(6) &$-$1.05(41)&   0.0077(31) \\
    4  &  16  &   35  &  16  & 1.017(5) &$-$2.37(55)&   0.0155(36) \\
    5  &  15  &  117  &  10  & 0.974(4) &$-$7.10(70)&   0.0430(43) \\
\end{tabular}
\caption{The number of data, $\chi^2$, and the calculated absolute
         normalizations $N_{abs}$ of the individual Uppsala data sets.
         Also, the $\chi^2(sl)$ and parameters $a$ and $b$ of the
         straight-line approximations.}
\label{tab:5}
\end{table}

\begin{references}
\bibitem{Error} Despite their awareness~\cite{Ols95} of this large
                discrepancy in the slopes, the authors of Ref.~\cite{Rah98}
                state: ``The agreement
                in the overlapping regions is generally very good.''
\end{references}
\end{document}